\documentstyle[11pt,gh2001-asp,twoside,epsf]{article}
\def\edcomment#1{\iffalse\marginpar{\raggedright\sl#1\/}\else\relax\fi}
\reversemarginpar

\def\magsqarcsec{mag$/$\raisebox{-0.4ex}{\hbox{$\Box^{\prime\prime}$}\,}}
\def\cf{cf.~}
\def\eg{e.g.~}
\def\eq{\!=\!}
\def\s0{S$0$}
\def\mubr{\hbox{$\mu_{\rm br}$}\,}
\def\rbrdhin{\hbox{\rbr$/$\hin}}
\def\hin{\hbox{$h_{\rm in}$}\,}
\def\rbr{\hbox{$R_{\rm br}$}\,}
\def\rco{\hbox{$R_{\rm co}$}\,}
\def\rcodh{\hbox{$R_{\rm co}/h_{\rm co}$}\,}
\def\rcodhinf{\hbox{$R_{\rm co}/h_{\infty}$}\,}
\def\hco{\hbox{$h_{\rm co}$}\,}
\def\vrot{\hbox{$v_{\rm rot}$}\,}

\def\halpha{$\mathrm{H}$\@{\sc$\alpha$}\,}
\def\hout{\hbox{$h_{\rm out}$}\,}
\def\hinf{\hbox{$h_{\infty \,}$}\,}
\begin{document}
\title{Radial structure of galactic stellar disks}
\author{M. Pohlen, R.-J. Dettmar, R. L\"utticke, \& G. Aronica}
\affil{Astronomisches Institut, Ruhr-Universit\"at Bochum \newline
Universit\"atsstr. 150, D-44780 Bochum, Germany \\}
\begin{abstract}
We present the results from our deep optical imaging survey 
\mbox{$\mu^{\rm V}_{\rm lim}\!\approx\!26-27$\,\magsqarcsec} 
of a morphologically 
selected sample of 72 edge-on disk galaxies. 
The question of the global structure of galactic stellar disks, especially 
the radial surface brightness profile at large galactocentric distances, 
is addressed.
We find that typical radial profiles are better described 
by a two-slope exponential profile ---characterised by an inner 
and outer scalelength separated at a 
break radius--- rather than a sharply-truncated exponential model.
Results are given for three face-on equivalents, serving as the 
crucial test to assure the findings for the edge-on sample without 
possible geometrical line-of-sight effects. 
\end{abstract}
\section{Introduction}
Van der Kruit (1979) initially found that the outer parts of disks of 
spiral galaxies do not retain their exponential light distribution 
to the observed faint levels, but rather show sharp edges.
For three nearby edge-on galaxies (NGC\,4244,\index{object, NGC 4244} 
NGC\,4565,\index{object, NGC 4565} NGC\,5907\index{object, NGC 5907}) 
he derived that the typical radial scalelength $h$ steepens from 5\,kpc 
to about 1.6\,kpc at the edge of the disk. 
The existence of these truncations, which are already visible in contour maps 
of edge-on and even of some face-on galaxies, is now well accepted 
(Pohlen 2001), but no unique physical interpretation is given
to describe this observational phenomenon.  
The proposed explanations span a rather wide range of 
possibilities. Van der Kruit (1987) deduced a connection to the 
galaxy formation process describing the truncations as remnants from 
the early collapse. Ferguson \& Clark (2001), for example,  
proposed an evolutionary scenario represented by the viscous disk 
evolution models. And Kennicutt (1989) suggested a ---probably less 
striking--- star-formation threshold.
Up to now the applied characteristic parameter for comparing 
the observational results of different studies is the distance 
independent ratio of a truncation radius $R_{\rm t}$ to a measured
radial scalelength $h$.
Van der Kruit \& Searle (1982) found for their sample of seven galaxies a 
value of $R_{\rm t}/h\eq4.2 \pm 0.6$, whereas Pohlen, Dettmar, \& L\"utticke 
(2000a) derived a significantly smaller one of $R_{\rm t}/h\eq2.9 \pm 0.7$ 
for their CCD survey of 30 galaxies.
\section{Our new edge-on sample}
To explain these different $R_{\rm t}/h$ values, we have improved 
the rather inhomogeneous sample of Pohlen et al.~(2000b).
The resulting morphologically selected sample contains 72 
galaxies of high data quality.
They are selected to be edge-on, undisturbed, and similar to 
'well-behaved' disk-prototypical cases such as NGC\,4565 
\index{object, NGC 4565} and IC\,2531.\index{object, IC 2531}
Thereby we want to assure that we are able to consistently fit our 
simple one-component disk model.
We have chosen S0-Sd galaxies (mainly Sb-Sc) and have obtained 
deep optical imaging in at least one filter, reaching 
a limiting surface brightness of $\mu^{1\sigma}_{\rm lim} \approx\!26-27$ 
V-\magsqarcsec or $\mu^{1\sigma}_{\rm lim} \approx\!25-26$ R-\magsqarcsec.
\begin{figure}
\plotfiddle{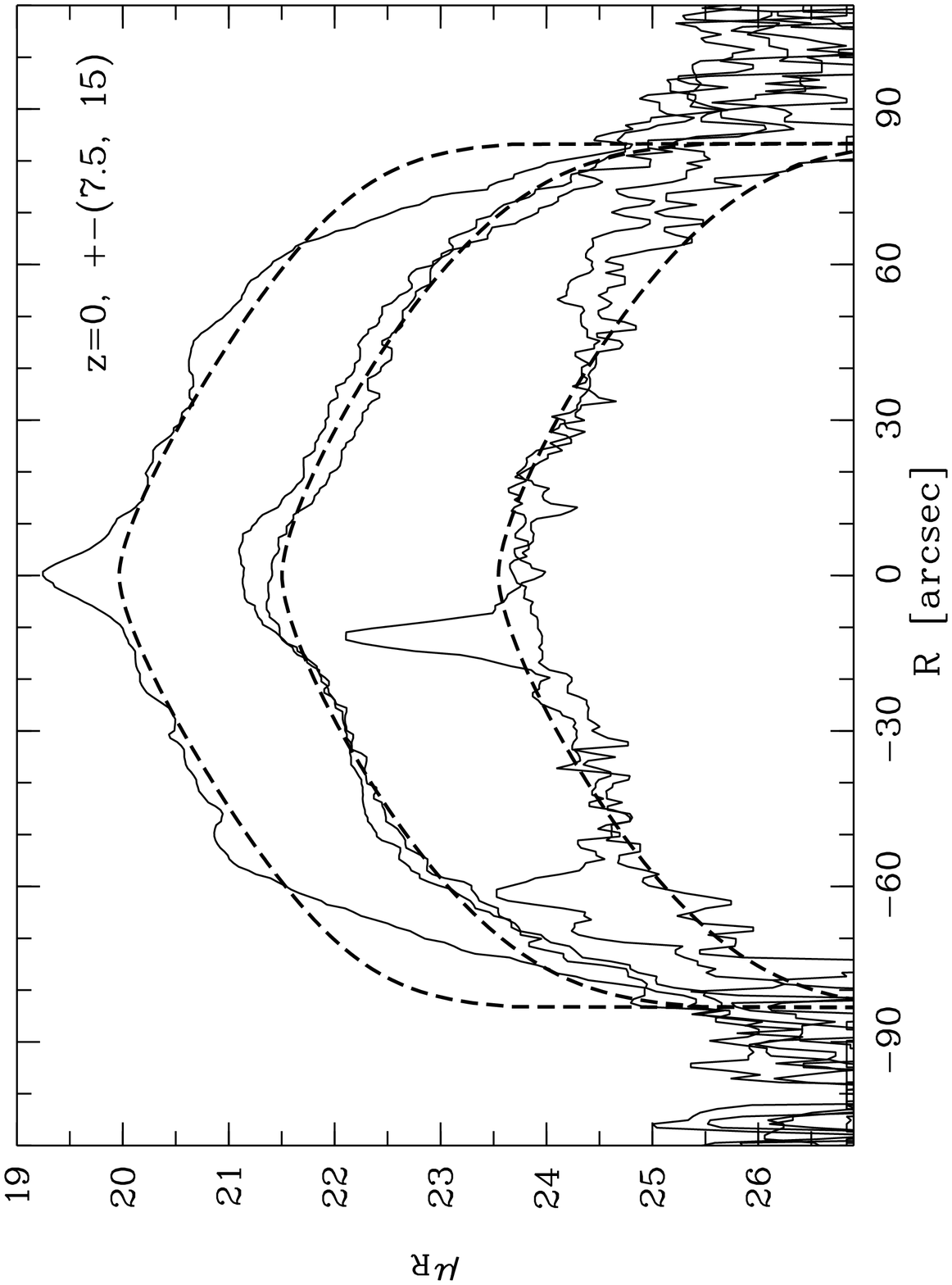}{1.5cm}{270}{23}{23}{-220}{71}
\plotfiddle{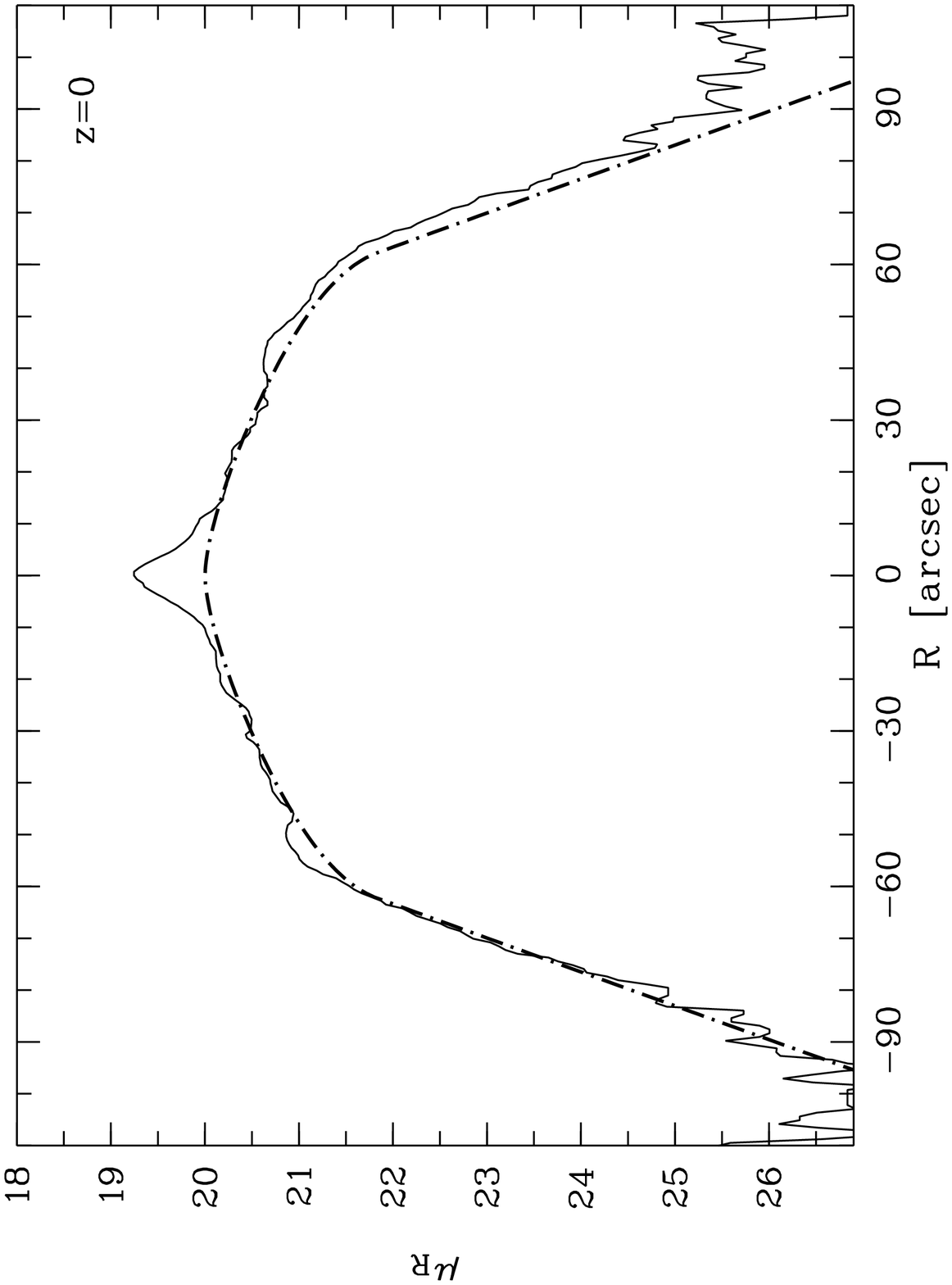}{1.5cm}{270}{23}{23}{0}{127}
\caption{Radial surface brightness profiles {\sl (solid lines)} of NGC\,522 {\sl (left panel)} \index{object, NGC 522} together with the best fitting sharply-truncated model {\sl (dashed line)} and the best fitting two-slope model {\sl (right panel)} .}
\end{figure}
We have fitted three different kind of models to the data (\cf Pohlen 2001): 
A sharply-truncated exponential model characterised by the cut-off 
radius \rco and the corresponding scalelength \hco. The cut-off radius 
is derived at the position where the profiles bend vertically into the noise.
An infinite-exponential model, which is only characterised by a 
scalelength \hinf and realised by fixing the cut-off radius 
to ten times the sharply-truncated scalelength, to address the problem 
of the a priori assumed sharp truncations.
And finally a two-slope or smoothly-truncated model,
characterised by a break radius \rbr and an inner \hin and an outer \hout
scalelength, which is obtained by fitting two 1D exponentials to the profiles.
\section{Results}
A surprising result is that sample galaxies, although morphologically 
selected to look as similar as our simple model disk, frequently show 
significant deviations from the input model.
Some objects exhibit a huge low surface brightness envelope 
(\eg ESO\,572-044,\index{object, ESO 572-044}) or a rather perturbed 
outer component (\eg UGCA\,250\index{object, UGCA 250}).
In another case (\eg ESO\,443-042\index{object, ESO 443-042}) a strong 
bar hampers a reasonable model fitting.
We find that three galaxies (NGC\,3390,\index{object, NGC 3390} NGC\,3717,
\index{object, NGC 3717} NGC\,4696C\index{object, NGC 4696C}), classified 
as Sb, do not show a truncation feature in their radial surface brightness 
profiles, but rather an \s0-like outer component.
For the remaining galaxies we derive a mean value of $\rcodh\eq3.5\pm0.8$
and confirm the suspected coupling of the two parameters, 
cut-off radius and associated scalelength, for the sharply-truncated model.
This is, in combination to a slightly different way in identifying 
the cut-off radius, the reason for the significantly deviant results 
of van der Kruit \& Searle (1982) and Pohlen et al.~(2000a).
The main result, however, is that most galaxies ($>\!60$\%) are best fitted 
with a two-slope or smoothly-truncated model.
Fig.~1 clearly shows that the best fitting sharply-truncated model
does not fit well for the inner profiles, which exhibit more likely
a two-slope behaviour.
The second slope is well described by another exponential decline.
We derive for the distance independent ratio of the break radius 
to the inner scalelength a value of $\rbrdhin\eq2.5\pm0.8$ (\cf Fig.~2)
with mean values of 7.6\,kpc for the inner and 1.9\,kpc for the 
outer scalelength.
The mean extrapolated surface brightness of this break on the major axis 
is $\mubr\eq23.4\pm0.6$\,V-\magsqarcsec and 
$\mubr\eq22.6\pm0.6$\,R-\magsqarcsec.
Additionally, 35\% of the galaxies are also well fitted at higher 
$z$-profiles with the two-slope structure.
We want to emphasise that the profiles show a rather sharp break, 
but are not sharply truncated, implying that beyond the break we still
find a remaining disk.
These two-slope profiles could not be produced by any conventional dust 
distribution as simulations have shown.
The various characteristic parameters (\eg\rcodh, \rcodhinf, \rbrdhin) do 
not correlate with the Hubble type, whereas plotting \rco and \rbr in linear 
units versus the maximum rotation velocity reveals an unusual 
distribution (\cf Fig.~2).
A general trend is expected since the faster the galaxy rotates 
the more mass it has and therefore its size will also be larger.
However, there is a surprisingly sharp limit in the size-velocity relation 
apparent. 
Above a well defined diagonal line no galaxies are found, 
whereas there is no similar lower boundary for the 
expected diagonal.
This would imply the existence of a maximal possible size for a given 
rotational velocity. 
\begin{figure}
\plotfiddle{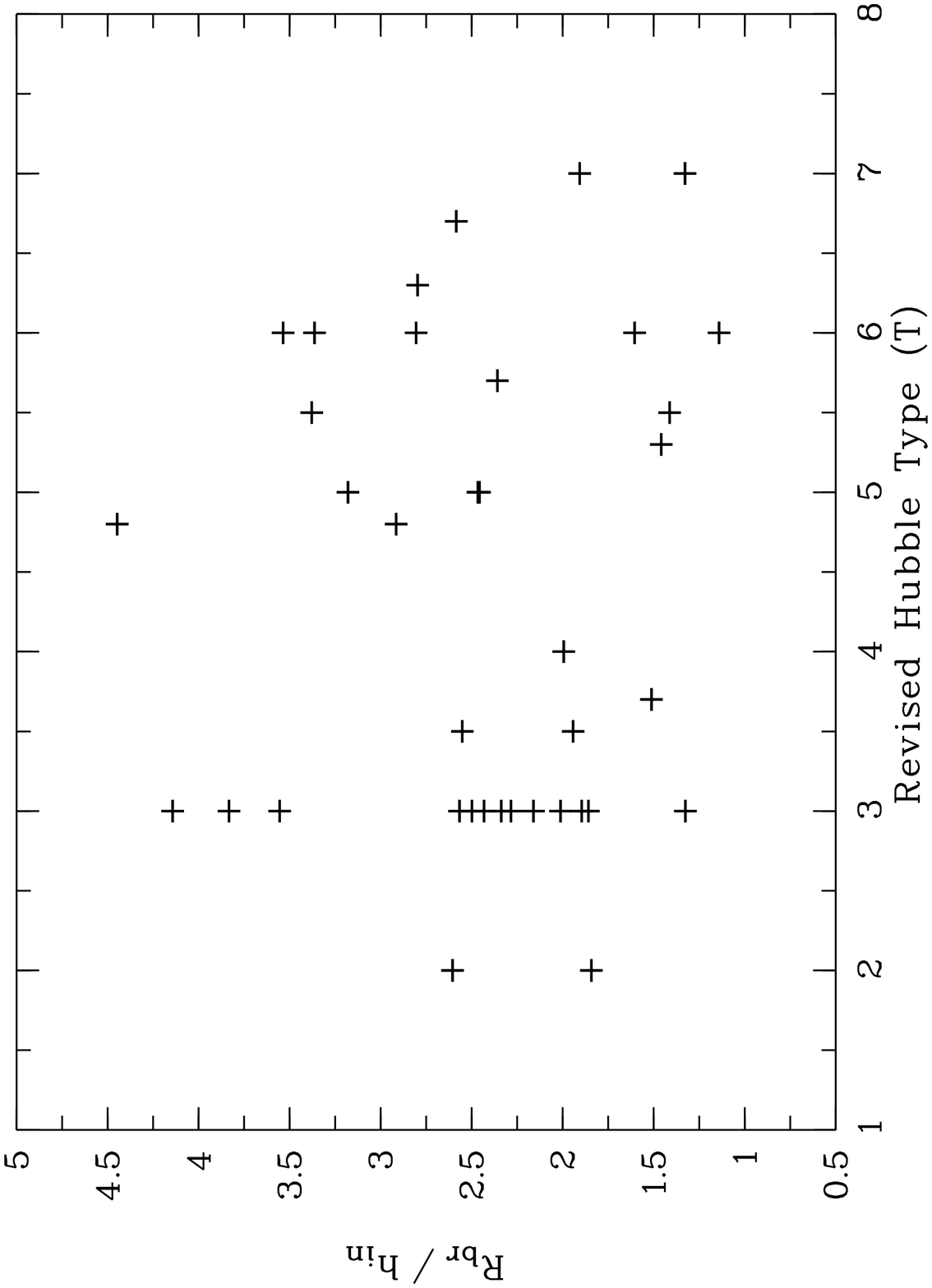}{1.5cm}{270}{23}{23}{-220}{71}
\plotfiddle{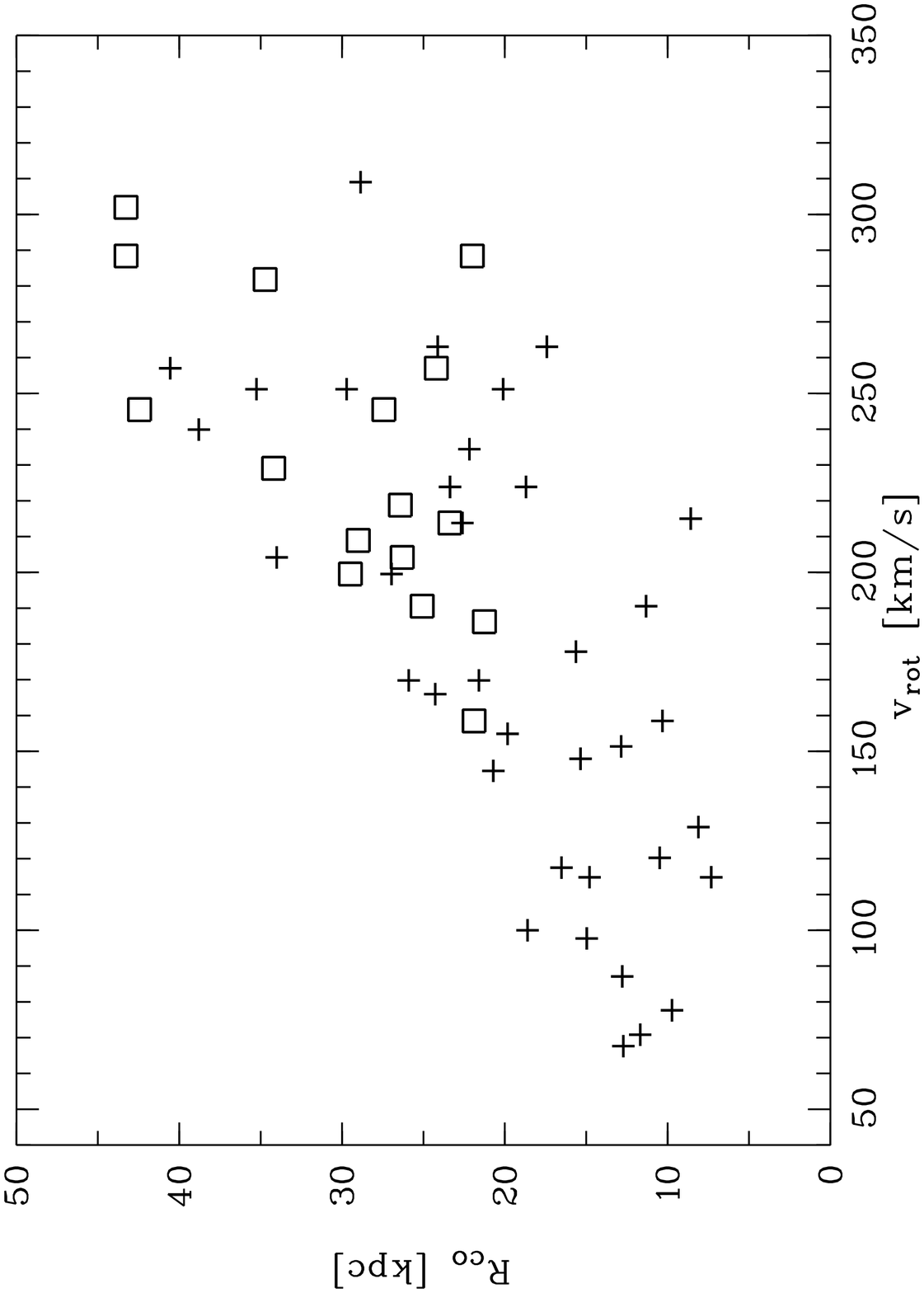}{1.5cm}{270}{23}{23}{0}{127}
\caption{{\sl Left panel:} Distribution of the distance independent ratio of \rbr to \hin with Hubble type $T$ {\sl (37 galaxies)}.  {\sl Right panel:} The distribution of cut-off radii versus maximum rotation velocity \vrot above {\sl square} and below {\sl cross} $D\eq50$\,Mpc {\sl (52 galaxies)}. \vspace*{-0.0cm}}
\end{figure}
The crucial experiment to confirm this two-slope structure is 
to observe the same behaviour for face-on galaxies. 
We have used the 2.2\,m telescope at Calar Alto with 
CAFOS and obtained images ($t_{\rm exp}\!\approx\!180$\,min) of three 
face-on galaxies, chosen to be as circular as possible and therefore 
intrinsically face-on and not of early type. 
Observations are made using an efficient rectangular R-band filter 
(R\"oser R, RR) achieving a reliable photometry down to
$\mu_{\rm RR}\eq28.0$\,\magsqarcsec, equivalent to 
$\mu_{\rm JR}\eq28.4$\,\magsqarcsec in Johnson R.
We find that the azimuthally averaged profiles as well as profiles 
from individual sectors exhibit a similar two-slope structure (\cf Fig.~3).
\begin{figure}
\plotfiddle{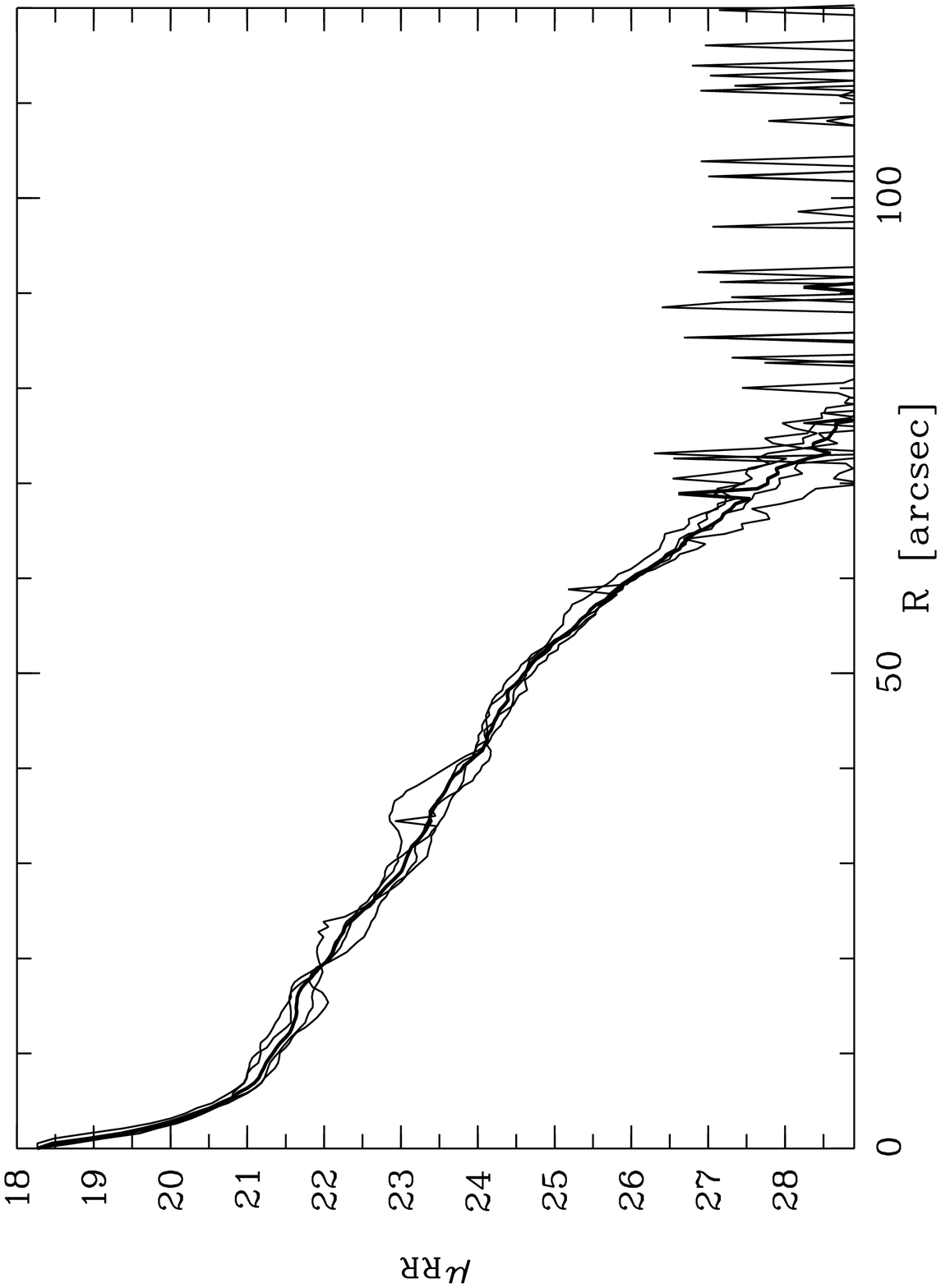}{1.5cm}{270}{23}{23}{-220}{71}
\plotfiddle{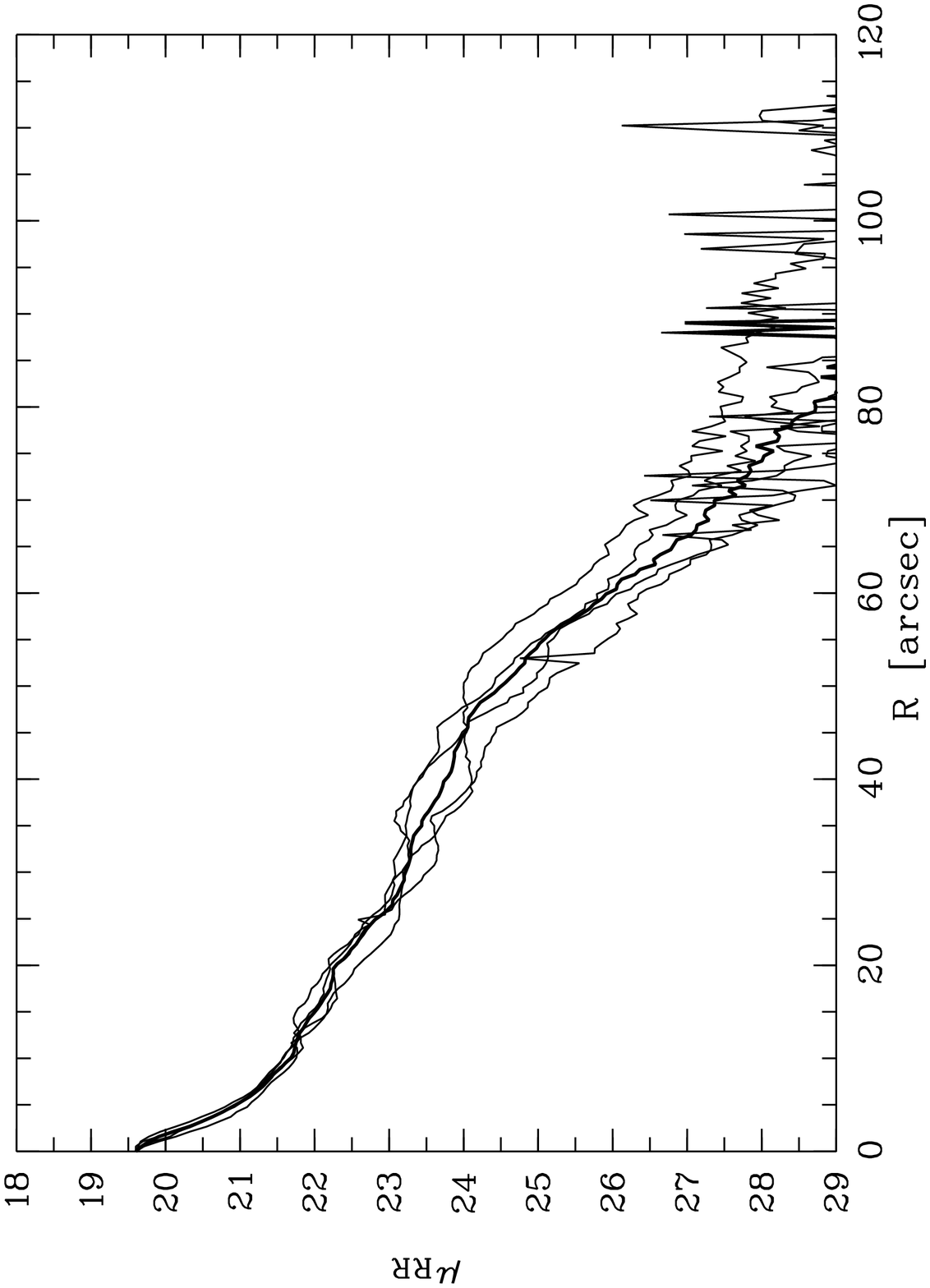}{1.5cm}{270}{23}{23}{0}{127}
\caption{Azimuthally averaged radial surface brightness profiles {\sl (solid lines)} of NGC\,5923 \index{object, NGC 5923}{\sl (left panel)} and UGC\,9837 \index{object, UGC 9837}{\sl (right panel)}. Plotted are the four 90\deg-segments {\sl (thin lines)} and the 360\deg {\sl (thick line)} combined profile. \vspace*{-0.0cm}}
\end{figure}
The break radius occurs at the same surface brightness level 
($\mubr\eq24.9\pm0.6$\,JR-\magsqarcsec) as compared to the edge-on case, 
if the line of sight integration is taken into account.
However, we derive a mean value for the ratio of the break radius to the 
inner scalelength $\rbrdhin\eq3.9\pm0.7$ (individually: 4.3, 4.2, and 3.1) 
for the three galaxies.
This does not fit well to the edge-on result and we probably still have
a problem to find a good scalelength comparison for the edge-on 
and face-on case.
However, the neglected dust and the applied 1D fitting 
of the individual scalelengths in the edge-on case both tend
to increase the measured scalelength and therefore decrease \rbrdhin.
\section{Outlook}
After this extensive optical imaging campaign the physical nature of 
disk truncations is still unknown and will be approached in the
next step.
We are still lacking of a detailed analysis to prove that these truncations
in the optical light profiles are also present in the mass distribution.
In addition, one has to check for possible environmental effects 
to address a tidal truncation scenario.
We will observe the radial molecular gas distribution ---as the reservoir for 
star-formation--- of galaxies with known optical truncations to find a 
correlation between the truncations and star-formation.
A similar approach will be performed by a comparison study of the truncated 
optical profiles with \halpha-profiles tracing the actual star-formation. 

\begin{references}
\reference Ferguson, A.M.N., \& Clark, C.J. 2001, \mnras, 325, 781
\reference Kennicutt, R.C., Jr. 1989, \apj, 344, 685
\reference Pohlen, M., Dettmar, R.-J., \& L\"utticke, R. 2000a, \aap, 357, L1 
\reference Pohlen, M., Dettmar, R.-J., L\"utticke, R. et al. 2000b, \aaps, 144, 405
\reference Pohlen, M., 2001, PhD Thesis, Ruhr-University Bochum, Germany
\reference van der Kruit, P.C. 1979, \aaps, 38, 15
\reference van der Kruit, P.C. 1987,  \aap, 173, 59
\reference van der Kruit, P.C., \& Searle, L. 1982, \aap, 110, 61
\end{references}
\end{document}